\documentclass[useAMS,usenatbib]{mn2e}
\usepackage[latin2]{inputenc} 
\usepackage[pdftex]{graphicx} 
\usepackage{hyperref} 
\usepackage{amsmath} 
\usepackage{amssymb} 
\usepackage{ulem} 
\usepackage{pdflscape}
\usepackage{lipsum}
\usepackage{color,soul} 
\usepackage{multicol}
\usepackage{graphicx}
\usepackage[export]{adjustbox}
\usepackage{tabularx} 
\usepackage{layouts}
\graphicspath{{./figs/}}
\newcommand{\HI}{H\,{\sc i} }
\newcommand{\Htwo}{H$_{2}$ }
\newcommand{\HInospace}{H\,{\sc i}}
\newcommand{\MHI}{M$_{\rm HI}$ }

\newcommand{\Mstar}{M$_{\star}$ }

\newcommand{\Msol}{M$_{\odot}$}

\makeatletter
\newsavebox\myboxA
\newsavebox\myboxB
\newlength\mylenA

\newcommand*\xoverline[2][0.75]{%
    \sbox{\myboxA}{$\m@th#2$}%
    \setbox\myboxB\null
    \ht\myboxB=\ht\myboxA%
    \dp\myboxB=\dp\myboxA%
    \wd\myboxB=#1\wd\myboxA
    \sbox\myboxB{$\m@th\overline{\copy\myboxB}$}
    \setlength\mylenA{\the\wd\myboxA}
    \addtolength\mylenA{-\the\wd\myboxB}%
    \ifdim\wd\myboxB<\wd\myboxA%
       \rlap{\hskip 0.5\mylenA\usebox\myboxB}{\usebox\myboxA}%
    \else
        \hskip -0.5\mylenA\rlap{\usebox\myboxA}{\hskip 0.5\mylenA\usebox\myboxB}%
    \fi}
\makeatother

\newcolumntype{b}{X}
\newcolumntype{s}{>{\hsize=.5\hsize}X}

\title[Gas as a Driver of the M{\it Z}R]
{The role of atomic hydrogen in regulating the scatter of the mass-metallicity relation }
\author[Toby Brown et al.]
{Toby Brown,$^{1,2}$\thanks{thbrown@swin.edu.au}
Luca Cortese,$^2$ 
Barbara Catinella$^2$
and Virginia Kilborn$^1$
\\
$^1$Centre for Astrophysics and Supercomputing, Swinburne University of Technology, Hawthorn, VIC 3122, Australia\\
$^2$International Centre for Radio Astronomy Research, University of Western
Australia,\\ 35 Stirling Highway, Crawley, WA 6009, Australia\\}

\begin{document}
\date{Accepted 2017 September 20. Received 2017 September 20; in original form 2017 June 05}

\pagerange{\pageref{firstpage}--\pageref{lastpage}} \pubyear{2002}

\maketitle

\label{firstpage}

\begin{abstract}
In this paper, we stack neutral atomic hydrogen (\HInospace) spectra for 9,720 star forming galaxies along the mass-metallicity relation. The sample is selected according to stellar mass (10$^9 \leq$ M$_{\star}$/M$_{\odot}\leq$10$^{11}$) and redshift ($0.02 \leq z \leq 0.05$) from the overlap of the Sloan Digital Sky Survey and Arecibo Legacy Fast ALFA survey. We confirm and quantify the strong anti-correlation between \HI mass and gas-phase metallicity at fixed stellar mass. Furthermore, we show for the first time that the relationship between gas content and metallicity is consistent between different metallicity estimators, contrary to the weaker trends found with star formation which are known to depend on the observational techniques used to derive oxygen abundances and star formation rates. When interpreted in the context of theoretical work, this result supports a scenario where galaxies exist in an evolving equilibrium between gas, metallicity and star formation. The fact that deviations from this equilibrium are most strongly correlated with gas mass suggests that the scatter in the mass-metallicity relation is primarily driven by fluctuations in gas accretion.
\end{abstract}


\section{Introduction}
\label{sec:Introduction}
Theoretical work has long predicted that galaxy growth is to a large extent regulated by the balance of gas accretion against star formation, and the subsequent dilution or enrichment of metals \citep[e.g.][]{Tinsley1978,Koppen1999}. In recent years, with the addition of outflows - the ejection of gas and metals from the interstellar medium (ISM) via energetic or momentum-driven winds - to this picture, we have begun to see the emergence of a framework for galaxy evolution where galaxies exist in a slowly evolving equilibrium between gas inflow, galaxy-scale outflows and star formation \citep{Finlator2008,Oppenheimer2008,Oppenheimer2010,Dave2011b,Dave2012,Lilly2013,Christensen2016}.

Observationally, the most common probe of this equilibrium is the relationship between stellar mass and gas-phase metallicity known as the mass-metallicity relation \citep[M$Z$R;][]{Lequeux1979,Garnett2002,Tremonti2004,Kewley2008,Zahid2011}. The general sense of the M$Z$R is that metallicity, as traced by O/H, increases linearly with stellar mass up to M$_{\star}$/M$_{\odot}\sim10^{10.5}$, after which the gradient flattens. There have been a number of mechanisms invoked to explain the observed slope and normalisation of the M$Z$R, including: i) outflows driving enriched gas from galaxies with greater efficiency at low stellar masses, where the shallow depth of the potential well means that material is easily ejected \citep[e.g.][hereafter T04]{Tremonti2004}, ii) mass-dependent interplay between chemical enrichment and dilution, primarily driven by the correlation between outflow strength and stellar mass \citep[e.g.][]{Finlator2008}, iii) increased metallicity of (recycled) accreted gas at higher stellar masses \citep[e.g.][]{Brook2014,Ma2016}, iv) the redshift evolution of an empirical stellar mass limit, above which the abundance of metals begins to saturate \citep[e.g.][]{Zahid2013}, and v) an effect of cosmic downsizing, where star formation and therefore chemical enrichment occurs preferentially in high mass galaxies at early-times \citep[e.g.][]{Maiolino2008,Zahid2011}.

Although the M$Z$R can be considered a tight scaling relation ($\sigma\sim$0.05-0.2 dex), its scatter is generally found to be at least a factor of two larger than the uncertainties present in the metallicity estimates \citep[T04,][]{Zahid2012}. Even more importantly, this dispersion is strongly correlated with other galaxy properties \citep{Cooper2008,Peeples2009,Dave2011b}. These two statements are the motivation for a significant amount of theoretical and observational effort to understand the physical drivers of scatter in the M$Z$R. The most commonly explored secondary dependency is the (anti-)correlation between metallicity and current star formation rate (SFR) at fixed stellar mass, first observed in Sloan Digital Sky Survey \citep[SDSS;][]{York2000} galaxies by \citet{Ellison2008} and since established, to varying extent, using a range of samples and methods at different redshifts \citep{Hunt2012,LaraLopez2013b,Stott2013,Cullen2014,Nakajima2014,Maier2014,delosReyes2015,Salim2015}. Since physical explanations for the origin of the M$Z$R generally invoke a balance between enriched outflows and pristine or recycled gas inflow as an explanation for the mean relation, deviations from this equilibrium (e.g. due to star formation) can be used as a probe of these mechanisms \citep[][herafter M10]{LaraLopez2010,Mannucci2010}.

Despite such studies, open questions remain as to the extent and physical nature of the metallicity-star formation rate dependency ($Z$-SFR). Early results by M10 seemed to show an invariance of the mass-metallicity-SFR (M$_{\star}$-$Z$-SFR) relation with redshift, resulting in those authors dubbing it the ``fundamental metallicity relation'', or FMR. However, more recent work suggests that, although the qualitative sense of the FMR persists beyond the local Universe, the normalisation and strength of the trend evolves with redshift \citep{Brown2016,Ma2017}. A dependence of the $Z$-SFR relationship on stellar mass was proposed by \citet{Yates2012}, who find that star forming, low mass galaxies (M$_{\star}$/M$_{\odot} \leq 10^{10.2}$) are indeed metal-poor, however, above this threshold they see a reversal where systems with higher SFR are found to have higher metallicities. This inversion of the M$_{\star}$-$Z$-SFR relation as function of mass is driven by the authors' choice of the T04 metallicity estimates. Further investigation of this result by \citet{Salim2014} suggests that it is the high signal-to-noise cuts on multiple emission lines that drives the correlation between T04 metallicities and SFR at higher stellar masses. \citet{Sanchez2013} go so far as to suggest that the $Z$-SFR relationship is driven by the presence of observational biases in SDSS data (i.e. fibre aperture effect), although several works have for the most part ruled out this conclusion by performing their own detailed analysis of the SDSS M$Z$R, showing that the relationship between metallicity and star formation persists even when these uncertainties are accounted for  \citep{Andrews2013,Salim2014,Telford2016}. 

In addition to the M$_{\star}$-$Z$-SFR relationship, it is reasonable to expect that there also exists an observable connection between the metallicity and gas content. Indeed, theoretical work generally supports this notion as a natural consequence of the equilibrium between inflows, outflows and gas processing \citep{Dutton2010,Dave2011b,Dave2012,Lagos2016}. Despite its theoretical importance, the intrinsic faintness of atomic gas (\HInospace) emission, poor statistics and selection biases have historically made accounting for the effect of gas on the M$Z$R much more difficult than that of star formation and only recently have studies begun to produce observational evidence for the M$Z$R dependency on gas. \citet{Hughes2013} use \HI observations of $\sim$250 objects to show that the \HI mass of metal-rich systems is typically lower than their metal-poor counterparts, attributing this to the increased efficiency of the star formation process in more massive galaxies. Using an increased sample of $\sim$4000 \HI selected galaxies, the work by \citet{Bothwell2013} shows an anti-correlation between gas mass and metallicity at fixed stellar mass, using this to argue that the \HInospace-M$Z$R relationship is more fundamental than the observed dependence on star formation. \citet{LaraLopez2013a} also find that galaxies with high gas fractions are metal-poor compared to their gas-poor counterparts. Lastly, recent investigations by \citet{Bothwell2016a,Bothwell2016b} highlight the importance of \Htwo as a regulator of the M$_{\star}$-$Z$-SFR relationship due to the correlation between \Htwo and SFR surface densities known as the Kennicutt-Schmidt law \citep{Schmidt1959,Kennicutt1998a}.

Based upon the work outlined above, it appears that an observable physical connection between the stellar mass of a galaxy, its gas-phase metallicity, gas content and current SFR exists. However, the exact character of this relationship remains elusive and establishing the importance of \HI content in comparison with star formation in this picture has so far not been possible. The strength of the M$Z$R dependencies upon gas and SFR, and subsequently the extent which these can be attributed to physical processes rather than observational techniques also remain unclear.

If we are to properly interpret theoretical predictions and develop a comprehensive understanding of chemical enrichment and star formation, we require in-depth studies using statistical samples for which star formation and \HI information is available. We may then establish the {\it relative} importance of atomic gas content and star formation in regulating the chemical evolution of galaxies. In this paper we stack atomic gas spectra of $\sim$10,000 star forming galaxies in order to investigate the robustness of \HI mass, global and fibre SFR in determining the scatter in the M$Z$R using two independent abundance calibrations. In doing so, this is the first work to demonstrate the reliability and dominance of gas content in comparison to SFR in setting the metal content of galaxies at fixed stellar mass, irrespective of the metallicity estimator.

This paper is organised as follows: Section \ref{sec:SampleandStacking} describes our sample and the quantities used in this work along with a brief description of the \HI stacking technique. In Section \ref{sec:HI-MZR} we examine the secondary drivers of the M$Z$R, quantifying its dependency on gas and star formation for a variety of SFR and metallicity indicators. The relative strengths of M$Z$R dependencies are compared in Section \ref{sec:DeltaX_MZR}. Section \ref{sec:Zgas_Discussion} outlines our conclusions and discusses them in the context of previous work. For clarity and succinctness, the potential biases that may be present in our work and how, if possible, we account for them are considered in Appendix \ref{Ap:PotentialBiases}.

\section{The Sample and \HI Stacking}
\label{sec:SampleandStacking}
\subsection{Sample Description}
\label{sec:DataSample}
We use a volume-limited ($0.02 \leq z \leq 0.05$), stellar mass-selected (M$_{\star}$/\Msol$\geq$10$^9$) parent sample \citep[see][]{Brown2015}. Briefly, this contains 30,368 galaxies with accompanying \HI spectra that are extracted using SDSS DR7 \citep{Abazajian2009} coordinates from the Arecibo Legacy Fast ALFA \citep[ALFALFA;][]{Giovanelli2005a} survey.

For the subset used in this work, galaxies must be classified as star forming according to the \citet{Kauffmann2003c} cut on the \citet*{Baldwin1981} diagram. Stellar masses and optical spectral line measurements are obtained by cross-matching with the publicly available Max Planck Institute for Astrophysics-Johns Hopkins University reduction of SDSS DR7 spectra\footnote{\url{http://wwwmpa.mpa-garching.mpg.de/SDSS/DR7/raw_data.html}} \citep{Brinchmann2004}. The emission line fitting methodology used by the MPA-JHU group is described in T04. We correct all strong emission lines for reddening using the Balmer decrement method assuming the extinction curve of \citet{Cardelli1989}.

We require that each galaxy has two valid estimates of gas-phase metallicity as measured by the abundance of oxygen with respect to hydrogen, log (O/H) + 12. For the first, we follow the selection criteria outlined by M10, which in turn are based upon the average of the R$_{23}$ and N2 methods recommended by \citet{Maiolino2008}. Secondly, we select the median log (O/H) + 12 estimates that are provided by T04 for star forming galaxies in the MPA-JHU catalogue. The T04 method differs significantly from other metallicity estimates in that it uses Bayesian analysis of theoretical model fits to the continuum-subtracted spectra rather than strong line ratios calibrated to metallicity. A full description of this methodology is given in T04.

Our justification for choosing the M10 calibration is that it has previously been used to show a significant dependence of the M$Z$R upon SFR and \HI content \citep[M10,][]{Bothwell2013,Salim2014}. It is an important check of our analysis that we are able to qualitatively reproduce the M$_{\star}$-$Z$-SFR and, in the case of \citet{Bothwell2013}, M$_{\star}$-$Z$-\HI relationships found in these works. Similarly, the T04 estimate is a commonly used diagnostic of the SFR dependence of the M$Z$R within the literature, however, the results are often in apparent conflict with those that use the M10 calibration \citep{Yates2012,LaraLopez2013b}. Furthermore, the Bayesian spectral energy distribution (SED) fitting approach used in the estimation of T04 metallicity is completely distinct from the methodologies used to calculate M10 (and other strong line ratio) metallicities. These two calibrations therefore characterise the uncertain nature of the M$Z$R and its secondary dependencies, allowing for an effective comparison and estimation of bias in the context of previous work.

To ensure that the SDSS 3 arcsec fibre covers a significant fraction of each galaxy in their sample, M10 implement a redshift cut (0.07 $<z<$ 0.3) that does not overlap in redshift with our sample. Although we cannot apply this cut for the main analysis, in consideration of the potential biases we are able to reproduce the M10 M$Z$R using the full SDSS DR7 (without the accompanying \HI data) and applying their redshift cuts. We provide a discussion of aperture effects in Appendix \ref{Ap:PotentialBiases}.

For each galaxy, we require that two SFR estimates are available. The first of these is the total SFR and is extracted from the GALEX-SDSS-WISE Legacy Catalog \citep[GSWLC;][hereafter S16]{Salim2016}, a database of stellar masses, SFRs and dust attenuation properties for $\sim$700,000 SDSS DR10 galaxies out to $z\sim0.3$ \citep[][]{Ahn2014}. These properties are derived using Bayesian SED fitting to optical and ultraviolet (UV) fluxes. The second indicator is the SFR contained within the 3 arcsec fibre and is calculated directly from the H$\alpha$ line flux, where S/N(H$\alpha$) $>$ 3, following Equation 2 in \citet[][hereafter K98]{Kennicutt1998b}.

Lastly, we also only select galaxies with stellar masses in the range $10^9 \leq$ M$_{\star}/{\rm M}_{\odot}\leq10^{11}$. This mass cut removes a further 8 galaxies while reducing the range of stellar masses in our most massive bin by 0.4 dex. The final sample consists of  9,720 star forming galaxies with: i) \HI spectra, ii) M10 and T04 estimates of gas-phase abundance, and iii) S16 global and K98 fibre SFR indicators.

\subsection{\texorpdfstring{\HI}{HI} Spectral Stacking}
\label{sec:Stacking}
Sensitivity limitations of \HInospace-blind surveys such as ALFALFA mean that obtaining statistical samples of galaxies spanning the entire gas-poor to -rich regime remains impossible using only detections. \HI spectral stacking analysis is commonly used to overcome this issue by co-adding 21 cm spectra  selected from \HInospace-blind surveys according to optical position and redshift, irrespective of whether they are detections or not. In recent years, this technique has been employed to successfully quantify the gas content of galaxies in the local Universe as a function of galaxy properties and environment \citep{Fabello2011a,Gereb2013,Brown2015,Brown2017}. In \citet{Brown2015} and \citet{Brown2017}, the authors weight the 21 cm spectra in each stack by the stellar mass of their corresponding galaxy, yielding an average gas fraction. It is important to note that in this work we do not weight by stellar mass, instead we bin the M$Z$R along both axes and compute the average \HI mass of each 2D bin by stacking the \HI spectra of galaxies within the bin. 

One of the caveats of stacking is that individual \HI masses are not available and therefore their distribution is not known. Thus, uncertainties on the average \HI masses used in this work are estimated using a modified delete-a-group jackknife technique \citep{Kott2001}. Briefly, for each \HI mass in this work our routine computes a further 5 stacked \HI masses while iteratively discarding, without repetition, a random 20\% of the galaxies that went into the original stack. The error is then the standard deviation of these 5 jackknifed estimates. The removal of a random 20\% of galaxies allows a reasonable number of jackknifed measurements while maintaining sufficient statistics in each bin so that the stacked signal is detected. As with any other jackknife method, this provides a robust statistical measure of the reliability of the stacked average gas content, however, it is not indicative of the distribution of gas content. A full description of the stacking technique and jackknifed errors used in this work is provided in Section 3 of \citet{Brown2015}.

\section{\texorpdfstring{\HI}{HI} and Star Formation as Regulators of the Mass-Metallicity Relation}
\label{sec:HI-MZR}
\begin{figure*}
      \centering
     \includegraphics[width=\textwidth]{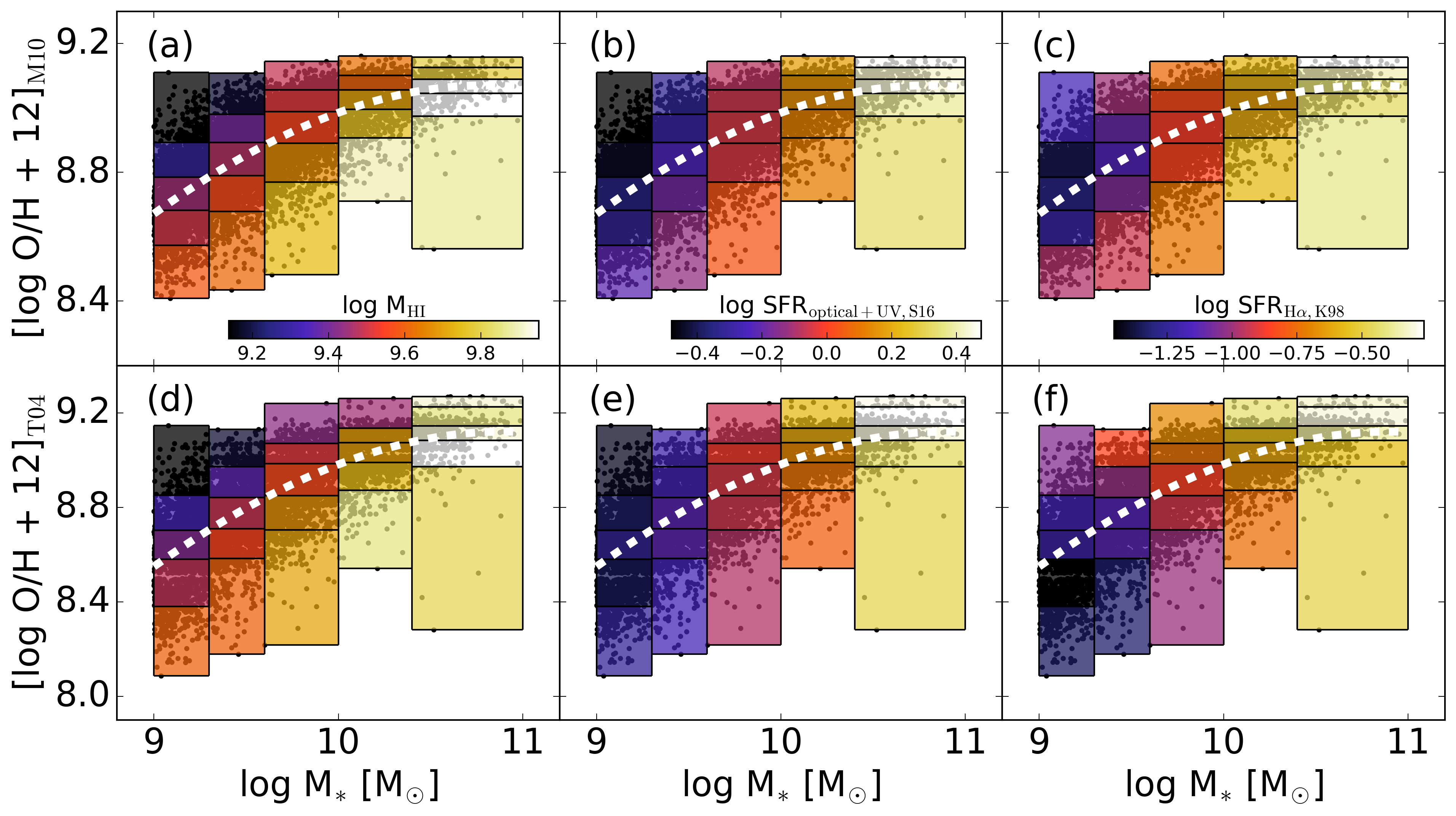}
  \caption{The M10 (top row) and T04 (bottom row) M$Z$Rs for galaxies in our sample. The 2D bins in the stellar mass-metallicity plane are shown by the boxes. Panels (a) and (d) respectively show the M10 and T04 M$Z$Rs colour-coded according to the average \HI mass in each bin. In (b) and (e), the M10 and T04 relations are coloured by average S16 total SFR while in (c) and (f) the colouring displays the average K98 fibre SFRs along each M$Z$R. \HI mass is measured in solar masses while estimates of SFR are quoted in M$_\odot$ yr$^{-1}$. The bin edges along the x-axis are log M$_{\star}$/M$_{\odot}$ = 9, 9.3, 9.6, 10, 10.4, 11. For each stellar mass bin, we also divide log (O/H) + 12 into bins that are $\pm 0.5\sigma, \pm1.5\sigma $ and $>1.5\sigma$ from the median M$Z$R. The corresponding colour bar for each column is given in the top panel. The white dotted lines are the best fit polynomials to the M10 and T04 relations.}
\label{fig:FMR_M10_T04}
\end{figure*}

\begin{figure*}
      \centering
     \includegraphics[width=\textwidth]{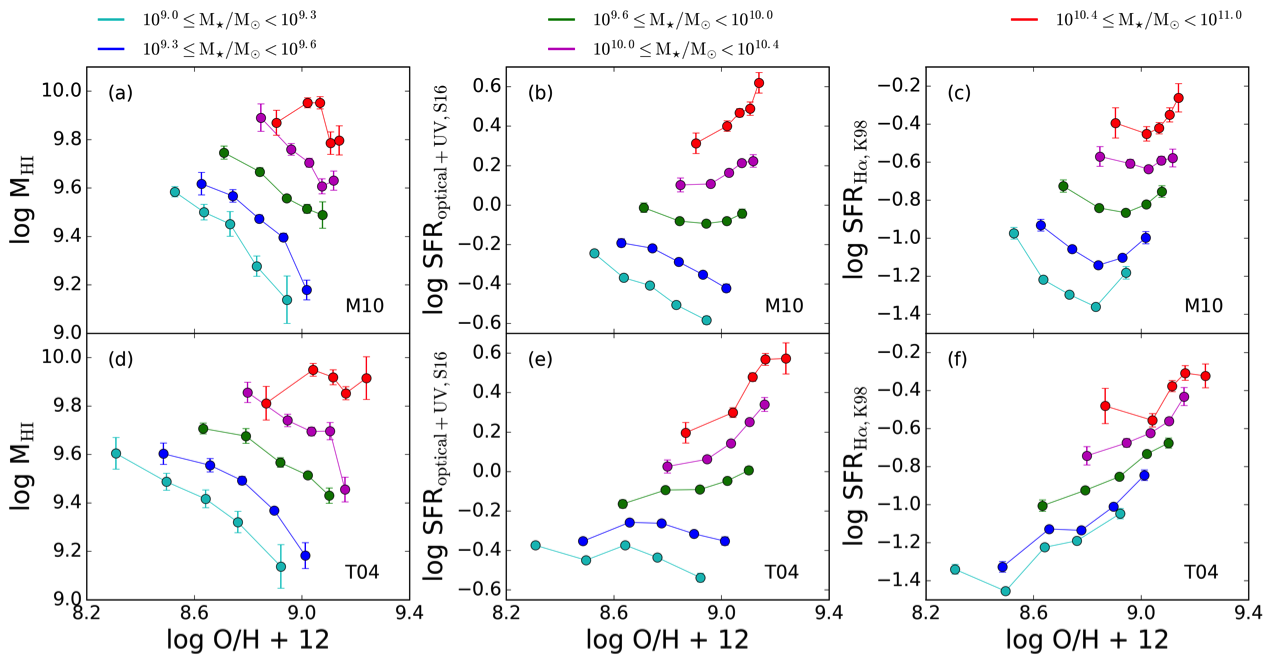}
  \caption{\HI mass (left column), S16 total SFR (middle column) and K98 fibre SFR (right column) as a function of log (O/H) + 12 for the M10 (top row) and T04 (bottom row) metallicity calibrations. \HI mass is in units of M$_\odot$ and SFRs are given in M$_\odot$ yr$^{-1}$. Y-axis values are the average quantity in each bin of metallicity at fixed stellar mass and are identical those used to colour the corresponding panels of Figure \ref{fig:FMR_M10_T04}. Stellar mass bins are given in the legend. Errors on average log \MHI are calculated using the Jackknife routine (see Section \ref{sec:Stacking}), while the standard error on the mean is used for both SFR estimates.}
\label{fig:X_Zgas_M10_T04}
\end{figure*}

In this section we look to identify the strongest driver of scatter in the M$Z$R. To do so, this work quantifies average \HI mass and SFR as a function of both M10 and T04 metallicity calibrations, at fixed stellar mass.

In Figure \ref{fig:FMR_M10_T04}, the black points in the background of each panel show the M10 (top row) and T04 (bottom row) M$Z$Rs for our sample. Their respective best fits, are denoted by the white dashed lines. For both M$Z$Rs, we recover the steep correlation between stellar mass and metallicity for galaxies with stellar mass below $\sim$10$^{10.5}$ M$_{\odot}$. Above this stellar mass, the gradient flattens until the correlation with mass disappears. As outlined in the introduction to this paper, this result has been noted by many observational and theoretical studies using a variety of samples and measurements of metallicity \citep[e.g. T04; ][M10]{DeLucia2004,deRossi2007,Kewley2008,Ellison2008,Finlator2008}. 

So that we may select sub-samples for stacking, we adopt two dimensional binning approach that is chosen to fully sample the dispersion in the M$Z$R while still having sufficient galaxies to stack and detect \HInospace. This involves dividing the sample first into 5 bins of stellar mass (limits are given in Figure \ref{fig:FMR_M10_T04} caption) and then again according to the metallicity percentiles (6.7\%, 30.9\%, 69.1\%, 93.3\%) corresponding to $\pm 0.5\sigma, \pm1.5\sigma $ and $>1.5\sigma$. In this way, we are able to compute the average \HI mass and SFR of galaxy populations in each bin at fixed stellar mass. As the M$Z$R changes depending on the metallicity calibration used, this method allows us to bin in a consistent manner using the M10 (a, b, c) and T04 (d, e, f) calibrations. The bins are illustrated by the boxes in each plot. Across each M$Z$R, the metallicity percentiles correspond to range in bin size of 0.04 to 0.7 dex with a median size of 0.1 dex. The number of galaxies per bin ranges from N=30 to N=1059, with a median of N=186. The width of the stellar mass bins increases as function of mass in order to properly sample the mass-dependent slope of the M$Z$Rs while also maintaining enough galaxies in each bin to obtain a detection when stacking. We stress that the highest mass bin is heavily affected by small number statistics, particularly below the M$Z$R. Thus, although we show the results for this range of stellar masses, caution must be used when interpreting any differences in trends with respect to the lower mass bins. Note that for consistency with the stacked \HI averages, we calculate the linear average of SFR in a bin and then take the logarithm.

In order to quantify star formation on a physical scale that is comparable to the \HI measurement, we use the GSWLC global SFRs from S16 that are derived via optical and UV SED fitting. This technique is known to provide a robust indication of global star formation properties \citep[][S16]{Brinchmann2004,Salim2007,daCunha2008} and, as such, has been used extensively within the field. In addition to global SFRs, we also want to measure the star formation over the same spatial scale as the metallicity is estimated (i.e. the star formation within the fibre aperture). Therefore, in parallel to the S16 SFRs explained above, we follow M10 and use the H$\alpha$ fibre SFRs calculated using the K98 prescription.

The left column of Figure \ref{fig:FMR_M10_T04} (a, d) shows the M$Z$Rs coloured according to the average \HI mass of galaxies within each 2D bin. For both the M10 (a) and T04 (d) calibrations, there is a strong anti-correlation of \HI mass with metallicity at fixed stellar mass, where galaxies below the M$Z$R have, on average, up to $\sim$0.6 dex less \HI mass than their high metallicity counterparts.

In the middle panels, boxes along the M10 (b) and T04 (e) relations are coloured according to their mean S16 SFR (SFR$_{\rm optical+UV, S16}$). In the low mass bins (M$_{\star}$/M$_{\odot}<$ 10$^{10}$), the top panel demonstrates an anti-correlation between metallicity and total SFR at fixed stellar mass, where galaxies with higher SFRs have lower metallicities and vice versa. The trend appears reversed for the more massive galaxies, here star formers appear to be metal-rich and gas-poor. We see for the T04-M$Z$R in panel (e), the low mass bins have a very weak correlation between metallicity and S16 total SFR, while the massive galaxies have a positive correlation.

On the right (c, f), we quantify the average K98 SFR in each box (SFR$_{\rm H\alpha, K98}$). In panel (c), the average fibre SFR echoes the anti-correlation with M10 metallicity exhibited by the total SFRs. However, in panel (f), when we look at the correlation between the average fibre SFR and T04 metallicities at fixed stellar mass we find a strong {\it positive} trend where galaxies that are star forming are also metal-rich. The relationship between metallicity and fibre SFRs is therefore contradictory for the two different abundance calibrations.

Figure \ref{fig:X_Zgas_M10_T04} quantifies the three-way dependency between the stellar mass, metallicity and \HI (or star formation) that are seen in the colour schemes of Figure \ref{fig:FMR_M10_T04} using a different projection. By plotting the average \HI mass, total SFR and fibre SFR, in the same bins as before, as function of the mean metallicity in each bin we show the $Z$-\MHI and $Z$-SFR relationships at fixed stellar mass for both M$Z$Rs. As shown above, \HI and metallicity are anti-correlated in each mass bin using both abundance calibrations while the $Z$-SFR relation depends on the metallicity estimator. However, using Figures \ref{fig:FMR_M10_T04} and \ref{fig:X_Zgas_M10_T04} for a direct comparison of i) each parameter's influence on metallicity (comparing horizontally) and ii) the sensitivity of different abundance calibrations to each parameter (comparing vertically) is not a straightforward task as the M$Z$Rs and y-axis are changing between panels.

\section{Identifying the Primary Driver of Scatter in the M\texorpdfstring{$Z$}{Z}R}
\label{sec:DeltaX_MZR}

\begin{figure*}
      \centering
     \includegraphics[width=\textwidth]{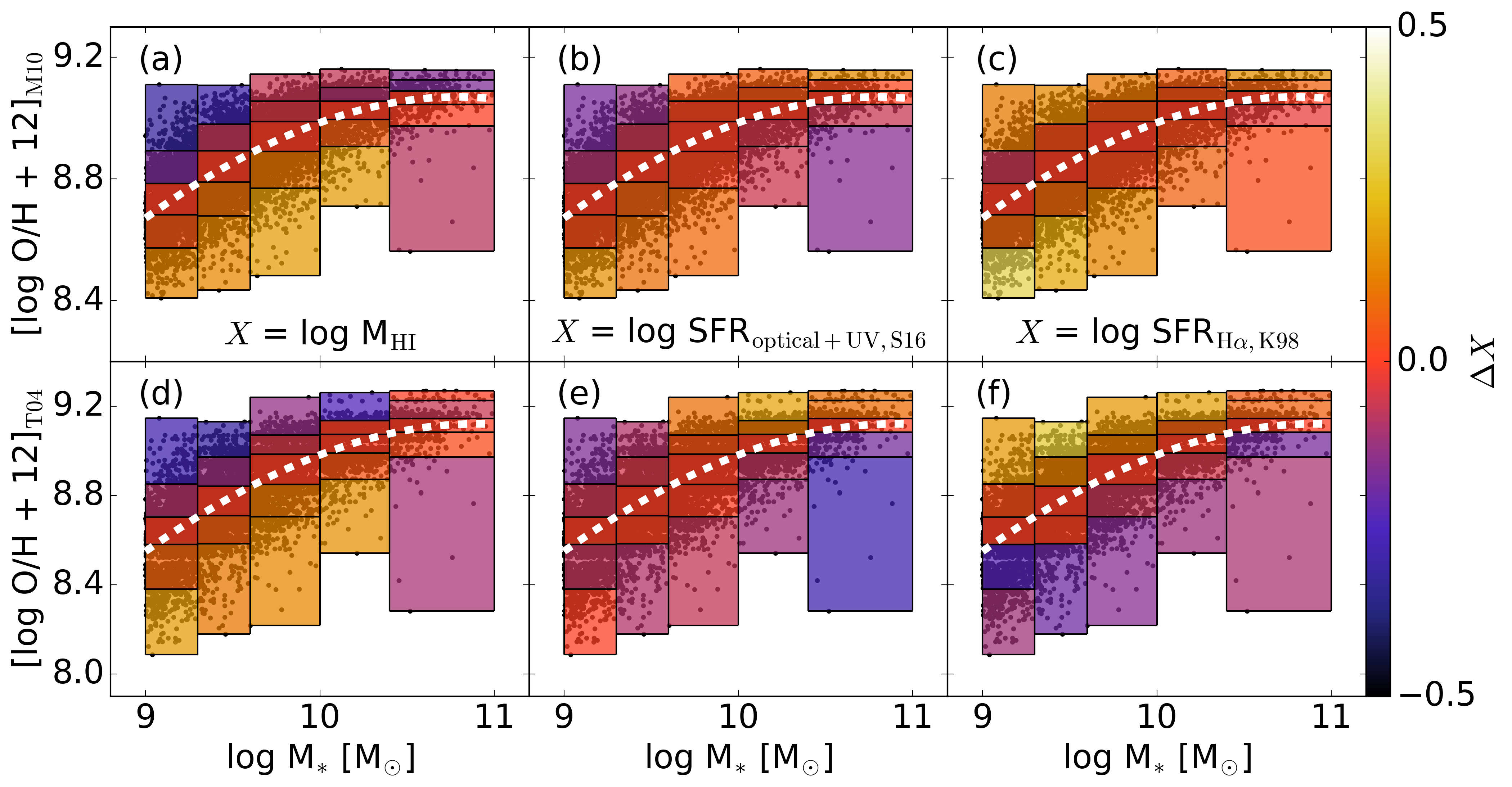}
  \caption{The M10 (top row) and T04 M$Z$Rs (bottom row). Each bin is coloured according to $\Delta X$, where $\Delta X = X - X_{MZR}$ (see Equation \ref{eq:DeltaX}) and, for each column,  $X$ is denoted in the top panel. The colour-coding is in dex and can be compared directly between panels. Black points, white dashed lines and 2D bin limits are identical to Figure \ref{fig:FMR_M10_T04}.}
\label{fig:FMR_dX_M10_T04}
\end{figure*}

\begin{figure*}
      \centering
     \includegraphics[height=0.43\textheight]{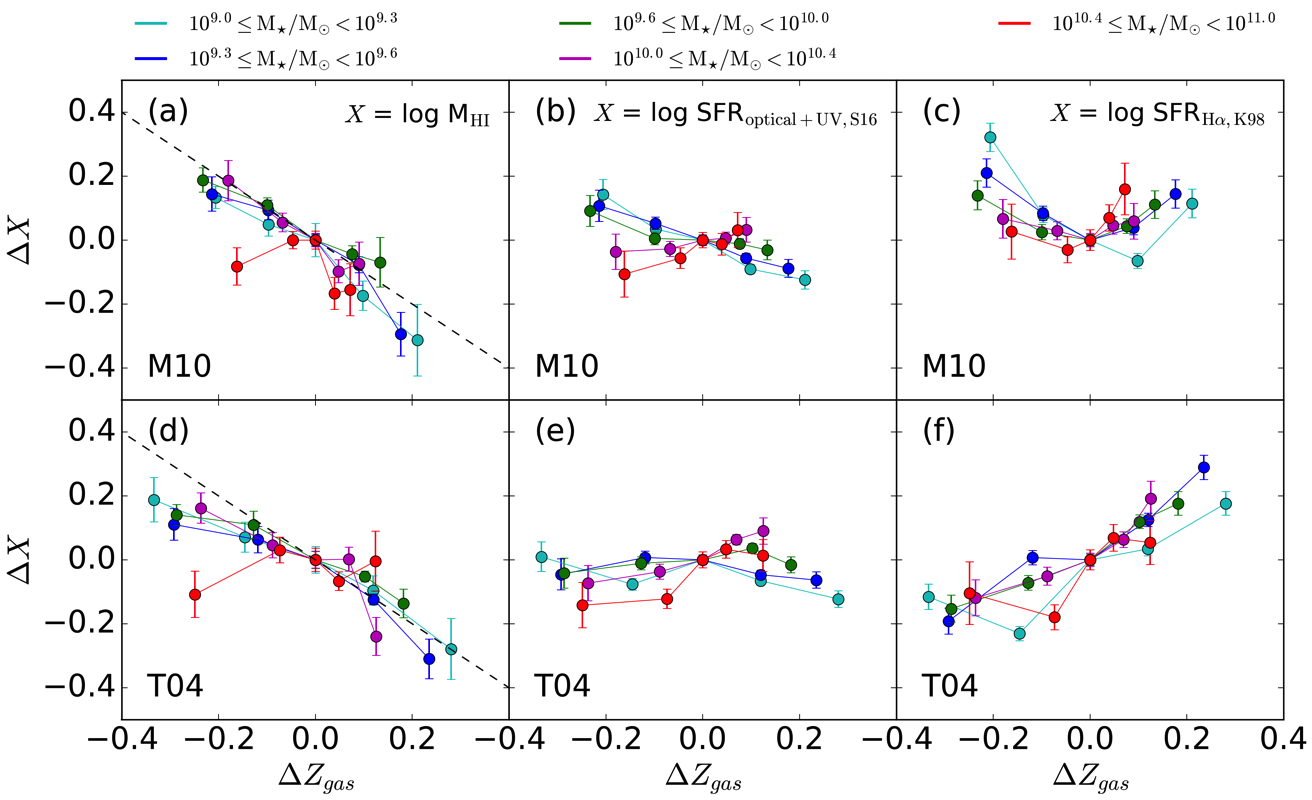}
    \caption{$\Delta X$ vs. $\Delta Z_{gas}$ for the M10 (top) and T04 (bottom) metallicity calibrations. Following the same pattern as Figures \ref{fig:FMR_M10_T04}, \ref{fig:X_Zgas_M10_T04} and \ref{fig:FMR_dX_M10_T04}, $X$ = log \MHI in left-hand column, log SFR$_{\rm optical+UV, S16}$ in the middle column and  log SFR$_{\rm H\alpha, K98}$ in the right-hand column. The stellar mass bins are identical to those figures and provided in the legend. For the left column, the dashed line shows the inverse one-to-one relation and error bars shown are DAGJK. In the middle and right columns we use the standard error on the mean.}
\label{fig:dX_dZgas_M10_T04}
\end{figure*}

To investigate observed tensions and compare the {\it relative} strength of the M$Z$R dependence upon \HI mass and SFR we use Figures \ref{fig:FMR_dX_M10_T04} and \ref{fig:dX_dZgas_M10_T04}. Firstly, this allows us to quantify the balance between gas, star formation and metallicity as a function of stellar mass (Figure \ref{fig:FMR_dX_M10_T04}) and, secondly, compare the influence of \HI mass, total SFR and fibre SFR in driving galaxies away from the equilibrium along the M$Z$R (Figure \ref{fig:dX_dZgas_M10_T04}).

In Figure \ref{fig:FMR_dX_M10_T04}, the top row shows the M10-M$Z$R while the bottom row uses the T04 metallicities. This time each box is coloured by the difference of a given parameter (M$_{\rm HI}$, SFR$_{\rm optical+UV, S16}$, SFR$_{\rm H\alpha, K98}$) from the value of that quantity on the M$Z$R at fixed stellar mass. This relative quantity, $\Delta X$, is therefore defined as:
\begin{equation}
\Delta X = \langle X \rangle - X_{MZR}
\label{eq:DeltaX}
\end{equation}
where $\langle X \rangle$ is the mean value of \HI mass (left column), S16 total SFR (middle) or K98 fibre SFR (right) in that bin. $X_{MZR}$ is the average value of the $X$ quantity within $\pm0.5\sigma$ of the M$Z$R in the same stellar mass bin. By definition, the value of $\Delta X$ along the M$Z$R is zero and, therefore, one can visualise these relative quantities as the offset in that property from the equilibrium population at fixed stellar mass. In other words, positive $\Delta X$ values are greater than and negative values less than the typical value of $X$ on the M$Z$R.

Since we are interested in the relative importance of \HI mass and star formation in regulating the position of galaxies on the M$Z$R, we suggest that $\Delta X$ is a more natural parameter for answering this question than the absolute value. The colour-coding across Figure \ref{fig:FMR_dX_M10_T04} is in dex, enabling direct comparison between all the panels.

At the same time, in Figure \ref{fig:dX_dZgas_M10_T04}, we quantify the influence of $\Delta X$ on the mean distance from the M$Z$R by plotting the offset in \HI mass ($\Delta$log M$_{\rm HI}$) and SFR ($\Delta$log SFR$_{\rm optical+UV, S16}$, $\Delta$log SFR$_{\rm H\alpha, K98}$) against the metallicity offset ($\Delta Z_{gas}$) for the M10 (top row) and T04 (bottom row) calibrations. Similarly to the other quantities, the offset in metallicity, $\Delta Z_{gas}$, is defined as:
\begin{equation}
\Delta Z_{gas} = \langle {\rm log (O/H) + 12}\rangle  - {\rm [log (O/H) + 12]}_{\rm MZR}
\label{eq:DeltaZgas}
\end{equation}
where $\langle$log (O/H) + 12$\rangle$ is the mean metallicity in a given box and [log (O/H) + 12]$_{\rm MZR}$ is the mean metallicity within $\pm0.5 \sigma$ of the M$Z$R at fixed stellar mass. Thus, $\Delta Z_{gas}$ is the average vertical scatter above (positive) or below (negative) the M$Z$R. The values of $\Delta X$ for each point are the same as those used to colour code the corresponding panels in Figure \ref{fig:FMR_dX_M10_T04}. Coloured lines show the $\Delta X-\Delta Z_{gas}$ relations in each stellar mass bin given in the legend. The relevant metallicity calibration is given in the bottom left corner of each panel.

Figure \ref{fig:FMR_dX_M10_T04}a shows the offset in \HI mass ($\Delta{\rm log\,M}_{\rm HI} = \langle{\rm log\,M}_{\rm HI}\rangle - {\rm log\,M}_{\rm HI, MZR}$) in each bin across the M10-M$Z$R. We see that galaxies that lie above the M10 relation are offset to lower gas masses than is typical for a galaxy on the M10-M$Z$R at that mass. Correspondingly, objects that are below the M10-M$Z$R are offset to higher \HI mass. Quantitatively, Figure \ref{fig:dX_dZgas_M10_T04} shows galaxies above the M10-M$Z$R (positive $\Delta Z_{gas}$) have up to $\sim$0.6 dex less \HI (negative $\Delta$log M$_{\rm HI}$) than their counterparts below the relation (negative $\Delta Z_{gas}$). Importantly, this anti-correlation is almost entirely independent of stellar mass.  The exception to this is at low metallicities in the highest mass bin, however, the large scatter and small number statistics in this bin mean that average \HI masses measured for this regime might not be reliable. The dynamic range in metallicity is smaller at higher stellar masses but the slope remains remarkably consistent between bins. Having said this, there is a hint that the $\Delta$log M$_{\rm HI}$-$\Delta Z_{gas}$ relationship is slightly non-linear, with the dependence of the M10 calibration on \HI appearing steeper above the M$Z$R than below when compared to the inverse one-to-one dashed line. 

Figure \ref{fig:FMR_dX_M10_T04}b presents the variations in total SFR across the same relation and, in the low mass bins (M$_{\star}$/M$_{\odot}\leq$ 10$^{10}$), the relationship between M10 metallicity and SFR$_{\rm optical+UV, S16}$ agrees with the \HInospace. Objects that are gas-poor and metal-rich are also, on average, more quiescent systems by $\sim$0.2 dex. However, it is only in the low mass regime that the anti-correlation of metallicity on total SFR holds true. At higher stellar mass, this trend reverses so that, on average, the more quiescent now lie $\sim$0.2 dex below the M10-M$Z$R. Panel (b) in Figure \ref{fig:dX_dZgas_M10_T04} shows the relationship between relative total SFRs and M$Z$R offset to be mass dependent. The three lowest mass bins have an anti-correlation between total SFR and M10 metallicities while the highest two mass bins exhibit a correlation. 

The anti-correlation between average fibre SFR and M10 metallicity is again present in the lower stellar mass bins, however relative differences in fibre SFR between star forming and quiescent populations are smaller  ($\sim$0.3 dex) than the dynamic range in \HI mass (a, $\sim$0.6 dex) and total SFR (b, $\sim$0.4 dex). Figure \ref{fig:dX_dZgas_M10_T04}c shows that the M10 metallicity trend for fibre SFRs differs above and below the M$Z$R. Where $\Delta Z_{gas}$ is negative, we see an anti-correlation between star formation and metallicity, while above M10-M$Z$R the relation is either flat (low masses) or weakly correlated (high masses). On the surface, the observed behaviour of the relationship between M10 metallicities and H$\alpha$ SFRs at high masses is in contradiction to the anti-correlation found by M10 in this regime. However, we note the large error bars on the K98 SFRs above the M10 M$Z$R at higher stellar masses. In addition, it is at these masses that the role of aperture effects in driving trends between metallicity and SFR is likely to be most apparent (see Appendix \ref{Ap:PotentialBiases}). We therefore caution against reading too much into the tension between our result and that of M10.

The bottom row of Figure \ref{fig:FMR_dX_M10_T04} shows the T04-M$Z$R. In panel (d), the trend between T04 metallicities and \HI content is found to be remarkably similar to the equivalent plot using M10 metallicities (a). Figure \ref{fig:dX_dZgas_M10_T04}d shows that the anti-correlation between relative \HI mass and T04-M$Z$R is mass independent and also remains non-linear, being steeper above the T04-M$Z$R than below. \HInospace-rich galaxies that are metal-poor again have $\sim$0.6 dex more \HI than gas-poor object, on average. In the low mass bins (M$_{\star}$/M$_{\odot}\leq$ 10$^{10}$) of Figure \ref{fig:FMR_dX_M10_T04}e, we see a lack of dynamic range in total SFR across the T04-M$Z$R ($\sim$0.2 dex ) compared the M10-M$Z$R ($\sim$0.4 dex, Figure \ref{fig:FMR_dX_M10_T04}b). At higher masses (M$_{\star}$/M$_{\odot}$$>$ 10$^{10}$), there is an increase in S16 SFR as function T04 metallicity at fixed stellar mass.

Figure \ref{fig:dX_dZgas_M10_T04}e shows the lack of trend between $\Delta Z_{gas}$ and total SFR for the low mass bins of the T04-M$Z$R where the relationship is flat. A positive correlation between the two parameters is present at higher stellar masses. When examining the scatter in the T04-M$Z$R using the fibre SFRs (\ref{fig:FMR_dX_M10_T04}f) the picture changes dramatically from what is seen with the total SFRs and with M10-M$Z$R. We find that, where in panel (\ref{fig:FMR_dX_M10_T04}c) there is a weak negative correlation between metallicity and SFR, there is now a positive correlation that persists independently of stellar mass. Figure \ref{fig:dX_dZgas_M10_T04}f quantifies the positive correlations between T04 metallicities and relative fibre SFR seen in Figure \ref{fig:FMR_dX_M10_T04}f. Interestingly, this relation is linear over the whole $\Delta Z_{gas}$ range and largely independent of stellar mass. This result appears inconsistent with the picture that is painted by the other metallicity-SFR relationships where there is an anti-correlation between star formation and metallicity at low stellar masses. 

To summarise, comparing the variation of colour across all 6 panels in Figure \ref{fig:FMR_dX_M10_T04} naturally leads to the conclusion that the M$_{\star}$-$Z$-SFR relationship is heavily dependent upon the combination of SFR indicator and metallicity calibration used as well as the stellar mass bin. On the other hand, the M$_{\star}$-$Z$-\HI relation is stronger, relatively stable across M10 and T04 calibrations and independent of stellar mass. The quantification of this result in Figure \ref{fig:dX_dZgas_M10_T04} suggests that the \HI content is a more reliable and physically motivated parameter than SFR for setting the metallicity at fixed stellar mass. In the next section we discuss this result and explore possible interpretations.

\section{Discussion and Conclusions}
\label{sec:Zgas_Discussion}
In this paper, we applied \HI spectral stacking to a sample of 9,720 galaxies with available metallicity, star formation and \HI information in order to determine whether gas content or star formation is the more physically motivated driver of scatter in the mass-metallicity relation.

The key conclusions can be can be summarised as follows:

\begin{enumerate}
  \item We confirm that there is an anti-correlation between the atomic gas content and gas-phase metallicity of galaxies at fixed stellar mass as previously found by \citet{Hughes2013}, \citet{Bothwell2013} and \citet{LaraLopez2013a}.
  \item The relationship between metallicity and gas content is consistent across the M10 and T04 M$Z$Rs as well as being largely independent of stellar mass. On the other hand, the dependency of metallicity on SFR is heavily reliant upon the choice of abundance calibration and star formation indicators used, and dependent upon the stellar mass of the system.
  \item Departures from the mean M$Z$R are more strongly correlated with \HI mass than either total or fibre SFR for both M10 and T04 abundance calibrations.
\end{enumerate}

The analysis in Section \ref{sec:HI-MZR} confirms that, at fixed stellar mass, increases in gas-phase metallicity above the equilibrium M$Z$R are correlated with decreases in \HI mass and, vice versa, deviations below the M$Z$R depend on increased gas content. Furthermore, we show the anti-correlation between gas and metals is significantly stronger than the dependency of metallicity on either total or fibre SFR. The observed M$_{\star}$-$Z$-\HI relationship also remains remarkably stable when using the M10 and T04 abundance calibrations while the character of the M$_{\star}$-$Z$-SFR relation changes depending on the metallicity and SFR indicators used. Discrepancies in the relationship between star formation and metals for the M10 and T04 calibrations have been reported by other studies \citep[e.g.][]{Yates2012,Salim2014}, however, this is the first work to demonstrate the reliability of gas content in setting the gas-phase oxygen abundance using both calibrations. Following this, we suggest that it is gas content, not star formation, that should be considered the {\it de facto} third parameter of the M$Z$R. 

The anti-correlation between metallicity and \HI content at fixed stellar mass found in our results can be explained by invoking changes in the rate at which gas is accreted. A rise in gas supply naturally leads to an increase in gas mass which, in turn, dilutes the metal abundance and boosts star formation. This is the situation we observe {\it below} the M$Z$R where we find galaxies to be more gas-rich and star forming. In the opposite scenario, a slowdown in infall rate means that processed gas is not replenished while ongoing star formation is simultaneously enriching the ISM. This acts to drive galaxies {\it above} the M$Z$R where, in the observations, we find them to be relatively gas-poor and quiescent. The fact that the M$_{\star}$-$Z$-\HI relationship also appears to be insensitive to stellar mass, suggests that this process drives departures from the M$Z$R effectively across the stellar mass range of our sample. The reduction in log (O/H) + 12 scatter as a function of stellar mass is consistent with a scenario where massive, gas-poor galaxies return to equilibrium faster \citep{Lagos2016}. While we do not rule out the influence of outflows in the dispersion of the M$Z$R, we note that it is difficult for this mechanism to produce the observed metallicity dependency on \HI content. Outflows acting to decrease metallicity would also eject gas from the disk, making the observed trend of metal-poor galaxies being the most gas rich unlikely. Thus, a more physically motivated picture is one where fluctuations in the rate of gas accretion are the primary driver of scatter in the M$Z$R at fixed stellar mass. The outflow rate, on the other hand, is strongly coupled to stellar mass of a galaxy and therefore it is possible this dictates the general form of the equilibrium M$Z$R relation.

This scenario is supported by a number of theoretical efforts, all of which are able to qualitatively reproduce our observed dependence of metallicity upon gas and star formation. \citet{Dave2012} and \citet{Lilly2013} used equilibrium models to connect inflow, outflow and star formation to galaxy metallicities. In both models, the deviations from the equilibrium at fixed stellar mass are primarily governed by the rate of gas accretion while the M$Z$R shape is governed by outflows. The analytic frameworks of \citet{Dayal2013} and \citet{Forbes2014} do not explicitly assume equilibrium yet their models also attribute the scatter in the M$Z$R to fluctuations in accretion. It should be noted that, in this case, the word ``accretion'' does not distinguish between the mechanisms of gas inflow along large-scale cosmic filaments \citep[e.g.][]{Keres2005,Dekel2009} and gas that is condensing onto the disk via the galactic fountain \citep[e.g.][]{Shapiro1976,Joung2006}.

\citet{Lagos2016} used the {\sc EAGLE} suite of cosmological hydrodynamical simulations \citep{Schaye2015} to investigate the metallicity of galaxies along the M$_{\star}$-SFR-gas fraction plane. The use of EAGLE means that \citet{Lagos2016} also relax the requirement for equilibrium as well as relying on fewer assumptions about the processing of gas than analytic models.
Using principle component analysis they determined that galaxy metallicity is most strongly correlated with gas fraction and suggest that galaxies self-regulate along this plane.

Lastly, we explore the non-linear nature of the $\Delta$\HInospace-$\Delta Z_{gas}$ relationship shown in Figures \ref{fig:dX_dZgas_M10_T04}a and \ref{fig:dX_dZgas_M10_T04}b. This result is intriguing because, if taken at face value, it suggests that regulation of metal content by \HI mass is more effective below the M$Z$R than above. The general sense of this statement is that the increase of gas supply above the equilibrium rate is more efficient at diluting metal content than a corresponding slowdown is at increasing metal concentration. The detailed modelling required to properly interpret such a result is considerably beyond the scope of this paper and will likely be the focus of future work. Here we simply note that the expected timeframe on which the ISM reacts to changes in the infall rate qualitatively supports this picture. Below the mean M$Z$R, an increase in gas supply would cause a decrease in metallicity and boost in star formation on a timescale that is of order a dynamical time \citep[$\sim$100 Myr;][]{Mo1998,Dave2011a}. Above the M$Z$R, the slowdown of inflow, consumption (or expulsion) of gas reservoirs and ongoing chemical enrichment can only occur on a timescale that is equivalent to M$_{\rm HI}$/SFR, known as the depletion time \citep[$\sim$1-10 Gyr;][]{Daddi2008}. Naively, this scenario would be expected to produce a skewed distribution of metallicities about the mean M$Z$R and, indeed, this is what we find in observations, where there is a tail in the log (O/H) + 12 distribution off to lower values at fixed stellar mass.

An alternative possibility is that the change in $\Delta$\HInospace-$\Delta Z_{gas}$ slope is due to the increasing contribution of \Htwo to the total gas mass as function of metallicity. It is straightforward to show that a correction of the gradient above the M$Z$R to match the gradient below requires an {\it increase} in \Htwo mass equivalent to $\sim$40\% of the \HI mass. This scenario seems plausible given the role of metallicity in the conversion of atomic to molecular hydrogen \citep{McKee2010}, however, the fact that such a large increase in the \Htwo mass at fixed stellar mass does not appear to be typical for star forming galaxies \citep[e.g.][]{Saintonge2016} is a significant caveat. For this reason, we find this explanation unlikely although further observations are required in order to test it fully.

To summarise, we quantify \HI mass, global and fibre SFR for $\sim$10,000 galaxies in two-dimensional bins across the M$Z$R, improving statistics beyond what has previously been possible. In doing so, we show that the observed relationship between \HI content and metallicity is more reliable and stronger than the relationship between metallicity and SFR. In other words, gas content rather than SFR should be considered the third parameter of the M$Z$R. We suggest that these results can be understood within the context of a model in which galaxies tend to grow in an equilibrium between gas content, metallicity and star formation where the scatter in the mass-metallicity relation is primarily driven by the fluctuations in the rate of gas supply. The results presented provides new and important observational constraints for theoretical models of galaxy evolution.

\section*{Acknowledgements}
The authors appreciate helpful comments and constructive suggestions from the referee. In particular, TB would like to thank Brent Groves for reading through and providing insightful comments on an early draft of this manuscript. We are also grateful to Claudia Lagos, Katinka Gereb and Steven Janowiecki for useful discussions when writing this paper. We acknowledge the work of the entire ALFALFA team in observing, flagging, and processing the ALFALFA data that this work makes use of. The ALFALFA team at Cornell is supported by NSF grants AST-0607007 and AST-1107390 and by the Brinson Foundation.

BC is the recipient of an Australian Research Council Future Fellowship (FT120100660). BC and LC acknowledge support from the Australian Research Council's Discovery Projects funding scheme (DP150101734).

Funding for SDSS-III has been provided by the Alfred P. Sloan Foundation, the Participating Institutions, the National Science Foundation, and the U.S. Department of Energy Office of Science. The SDSS-III web site is \url{http://www.sdss3.org/}.

SDSS-III is managed by the Astrophysical Research Consortium for the Participating Institutions of the SDSS-III Collaboration including the University of Arizona, the Brazilian Participation Group, Brookhaven National Laboratory, Carnegie Mellon University, University of Florida, the French Participation Group, the German Participation Group, Harvard University, the Instituto de Astrofisica de Canarias, the Michigan State/Notre Dame/JINA Participation Group, Johns Hopkins University, Lawrence Berkeley National Laboratory, Max Planck Institute for Astrophysics, Max Planck Institute for Extraterrestrial Physics, New Mexico State University, New York University, Ohio State University, Pennsylvania State University, University of Portsmouth, Princeton University, the Spanish Participation Group, University of Tokyo, University of Utah, Vanderbilt University, University of Virginia, University of Washington, and Yale University.

\bibliographystyle{mn2e}
\bibliography{refs}

\appendix
\section{A Discussion of Potential Biases}
\label{Ap:PotentialBiases}
\subsection{Aperture Effects}
\label{sec:Aperture_Effects}

\begin{figure*}
      \centering
     \includegraphics[width=\textwidth]{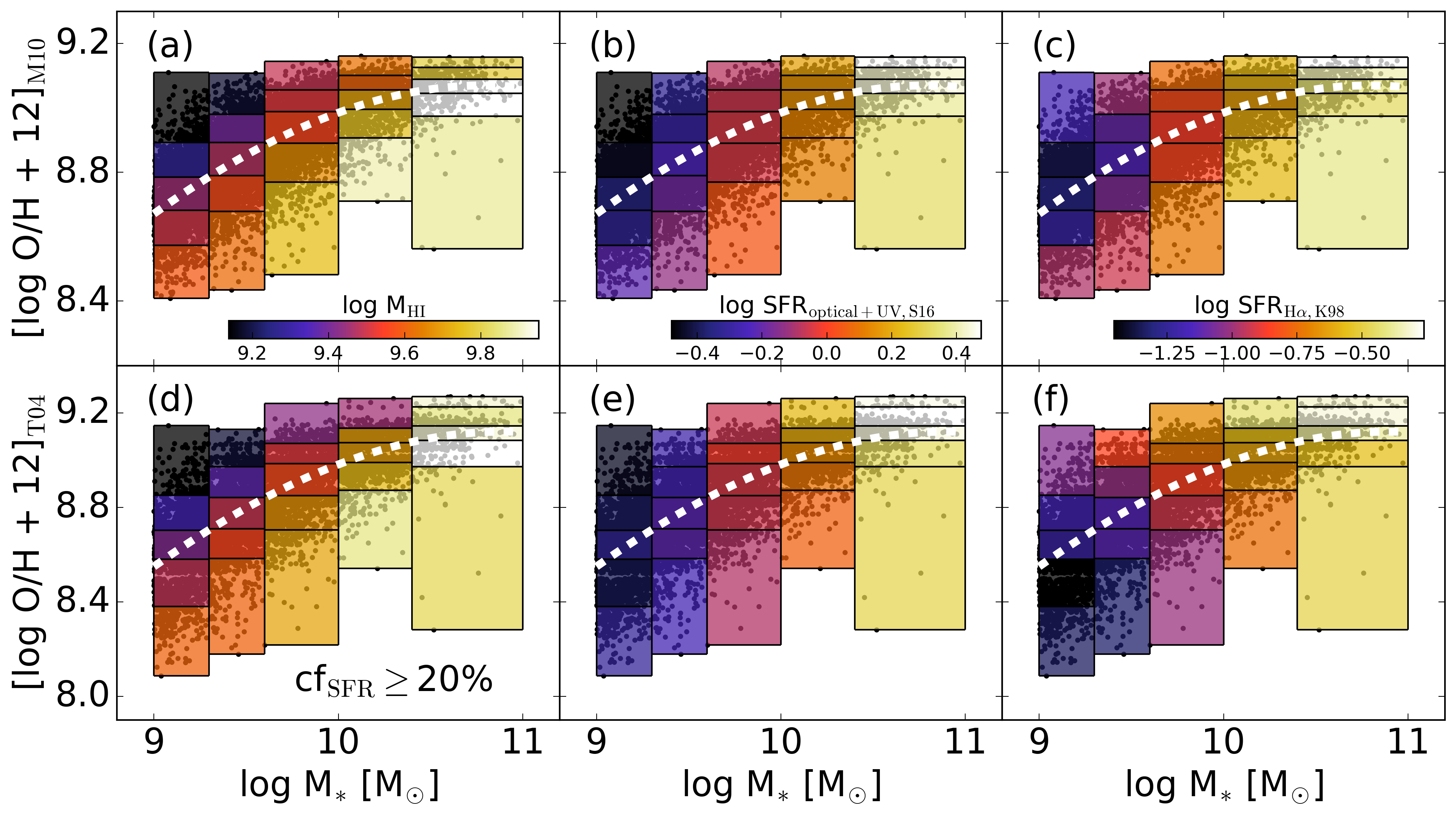}
  \caption{The same as Figure \ref{fig:FMR_M10_T04} using the sample of 5014 galaxies that have a star formation covering fraction (cf$_{\rm SFR}$, the ratio of fibre-to-total SFR) $\geq$20\%.}
\label{fig:FMR_M10_T04_sfr_covfrac}
\end{figure*}


We now consider the role that the 3 arcsec aperture of the SDSS fibre plays in driving the secondary dependencies of the M$Z$R seen in Section \ref{sec:HI-MZR}. It is well known that systematic gradients in the metallicity as a function of radius are present in most massive late-type galaxies with metallicities typically found to be higher in the centre than at the outskirts \citep{Zaritsky1994}. To avoid sampling only the central regions and therefore the abundance measurement being biased high, most studies using fibre spectroscopy tend to either set a lower redshift bound on their samples so that the fibre corresponds to a reasonable physical size \cite[e.g. M10,][]{Salim2014,Telford2016}, or use a slightly more sophisticated cut to select galaxies for which the covering fraction of the SDSS fibre is large \citep{Ellison2008,Kewley2008}. Previous work has found that a flux covering fraction (cf$_{flux}$, ratio of fibre-to-total flux) of 20\% is suitable for recovering fibre derived abundances that agree well with estimates of global metallicity for galaxies with M$_{\star}$/M$_{\odot}<$10$^{10}$ \citep[][]{Kewley2005,Kewley2008}. In their paper, \citet{Kewley2008} conclude that a covering fraction of $>$20\% may be required to avoid metallicity gradients in galaxies of M$_{\star}$/M$_{\odot}>$ 10$^{10}$. The 3 arcsec diameter of the SDSS fibre corresponds to between 1 and 3 kpc across the redshift range of our sample and is therefore typically smaller than the average galaxy size (the sample's mean radius is $\sim$3 kpc).

Unfortunately, the requirement of \HI data and large statistics for this work mean that such a cut is not an option for our analysis. Instead, we choose to check our results by ensuring they are not driven by SFR covering fraction. To do so, we repeat the analysis in Figure \ref{fig:FMR_M10_T04_sfr_covfrac} using a reduced sample of 5014 galaxies for which the ratio of fibre-to-total SFR (cf$_{\rm SFR}$) is more than 20\%. We use a SFR ratio instead of a flux ratio for this criterion because only $\sim$30\% of the sample has a cf$_{flux}$ over 20\%. Nevertheless, this cut is the same proxy used by \citet{Bothwell2013} and follows the same trends as the more conservative covering fraction estimates. In Figure \ref{fig:FMR_M10_T04_sfr_covfrac} we plot the M10 and T04 relations for the $\sim$5000 galaxies for which cf$_{\rm SFR}\,\geq$20\%. Encouragingly, each of the relationships between metallicity and \HI (or SFR) remains consistent with what we find in Section \ref{sec:HI-MZR}. We use this cut for all our galaxies in order to ensure we achieve an \HI detection in the high mass regime, increasing the cf$_{\rm SFR}$ cut higher than 20\%  would make this impossible.

While we cannot rule out that our results are in some part dependent on biases introduced by to metallicity gradients and the fibre aperture, particularly at larger masses (\Mstar $\geq$10$^{10}$ M$_{\odot}$), the fact that the trend persists after this cut suggests that the main conclusions of this paper are not likely to be driven by the aperture effects. This is supported conclusion is supported in more detailed  analysis of systematics within the M$Z$R trends \citep{Andrews2013,Salim2014,Telford2016}.

\subsection{Choice of Metallicity and Star Formation Rate Indicators}
\label{sec:Zgas_bias}
Given the significant practical and theoretical challenges present in the measurement of metallicities, determining the `best' calibration is not a trivial task. As such, the different calibrations and their respective advantages remain a subject of much interest within the field. An in-depth discussion on this topic is well beyond the scope of this paper and we therefore refer readers interested in a more thorough discourse to works of \citet{Kewley2008}, \citet{Andrews2013} and \citet{Salim2014}, and the references contained therein. For this work, our objective is simply to ensure that key results are not driven to a significant extent by our choice(s) of metallicity calibration and it is for this reason that we select two calibrations that `bracket' the range of disagreement found in the literature. 

We choose the \citet[][hereafter KD02]{Kewley2002} metallicity calibration as a sanity check for our results. Successfully reproducing the mean M$Z$R found in that work. However, the strict signal-to-noise emission line cuts in the KD02 method mean that there are only $\sim$5000 galaxies with valid KD02 metallicities. Unfortunately, these statistics are not sufficient for recovering the stacked \HI mass across the scatter of the M$Z$R. We note that for it is very difficult to imagine a scenario where either the choice of abundance calibration and/or aperture effects can result in a trend between gas and metallicity that is both strong and stable. Observations searching for the dependence of the M$Z$R on star formation have so far yielded vastly different results \citep[e.g. M10, ][]{Yates2012,Sanchez2013}. Considering this uncertainty, it is important to recognise that this is not the case for studies of the M$_{\star}$-$Z$-\MHI relationship. Using significantly different samples, each group investigating this dependence has found a qualitatively similar picture - gas content is anti-correlated with metallicity \citep{Hughes2013,Bothwell2013,LaraLopez2013a}.

To check that the choice of S16 star formation rates is not driving some of the trends we find, we repeat the analysis in Figures \ref{fig:FMR_M10_T04} to \ref{fig:dX_dZgas_M10_T04} using the global star formation rates from the MPA-JHU catalogue \citep{Brinchmann2004}. Since the dependency of the M$Z$R on the S16 SFRs is very similar to the trend with the MPA-JHU SFRs we do not present the comparison figures in this paper. The similarity between the M$Z$Rs dependence on the two total SFR indicators is not surprising given the strong correlation between these two estimates for star forming galaxies \citep[see][]{Salim2016}.
\end{document}